\documentclass[preprint,aps,amsmath,nofootinbib,tightenlines,floatfix]{revtex4}
\usepackage{graphicx,epsfig,amssymb}
\usepackage{multirow}

\begin{document}
\title{Dilepton Signals in the Inert Doublet Model}
\author{Ethan Dolle\footnote{edolle@physics.arizona.edu}, 
        Xinyu Miao\footnote{miao@physics.arizona.edu}, 
        Shufang Su\footnote{shufang@physics.arizona.edu}, and 
        Brooks Thomas\footnote{brooks@physics.arizona.edu}}
\affiliation{Department of Physics, University of Arizona, Tucson, AZ  85721 USA}

\begin{abstract}
The Inert Doublet Model is one of the simplest and most 
versatile scenarios for physics beyond the Standard Model.
In this work, we examine the prospects for detecting the
additional fields of this model at the LHC in 
the dilepton channel.  We investigate a wide variety of
theoretically- and phenomenologically-motivated benchmark scenarios,
and show that within regions of model parameter space in which
the dark-matter candidate is relatively light (between $40$ and $80$~GeV) 
and the mass splitting
between the neutral scalars is also roughly $40 - 80$~GeV, a 
signal at the $3\sigma$ to $12\sigma$ significance level can be observed with 
$100\mbox{ fb}^{-1}$ of integrated luminosity.  In addition, even if   
the mass splitting between the neutral scalars is larger than $M_Z$, 
a signal of more than $3\sigma$ can be observed as long as the mass of the
dark-matter candidate is around 40~GeV. 
\end{abstract}

\maketitle

\newcommand{\newc}{\newcommand}
\newc{\gsim}{\lower.7ex\hbox{$\;\stackrel{\textstyle>}{\sim}\;$}}
\newc{\lsim}{\lower.7ex\hbox{$\;\stackrel{\textstyle<}{\sim}\;$}}

\def\beq{\begin{equation}}
\def\eeq{\end{equation}}
\def\beqn{\begin{eqnarray}}
\def\eeqn{\end{eqnarray}}
\def\calM{{\cal M}}
\def\calV{{\cal V}}
\def\calF{{\cal F}}
\def\half{{\textstyle{1\over 2}}}
\def\quarter{{\textstyle{1\over 4}}}
\def\ie{{\it i.e.}\/}
\def\eg{{\it e.g.}\/}
\def\etc{{\it etc}.\/}


\def\inbar{\,\vrule height1.5ex width.4pt depth0pt}
\def\IR{\relax{\rm I\kern-.18em R}}
 \font\cmss=cmss10 \font\cmsss=cmss10 at 7pt
\def\IQ{\relax{\rm I\kern-.18em Q}}
\def\IZ{\relax\ifmmode\mathchoice
 {\hbox{\cmss Z\kern-.4em Z}}{\hbox{\cmss Z\kern-.4em Z}}
 {\lower.9pt\hbox{\cmsss Z\kern-.4em Z}}
 {\lower1.2pt\hbox{\cmsss Z\kern-.4em Z}}\else{\cmss Z\kern-.4em Z}\fi}




\section{Introduction\label{sec:Introduction}}

The Inert Doublet Model~\cite{FirstIDM} (IDM) is one of the 
simplest extensions of the Standard Model (SM), yet
it is also one of the most versatile.  This scenario, 
in which the usual SM fields are supplemented 
by a single scalar ${\rm SU}(2)$ doublet which does not contribute to
electroweak-symmetry breaking (EWSB) and couples to the 
gauge-boson sector but not the fermion sector, has a wealth of
potential phenomenological applications.  Perhaps the
most intriguing of these stems from the recent 
observation~\cite{BarbieriIDM} that the fields of this
additional scalar doublet can provide a positive contribution 
to the oblique $T$ parameter~\cite{STU} sufficient to 
render a SM Higgs mass of $m_h = 400 - 600$~GeV consistent with 
precision data~\cite{LEPBound}.  A host
of other potential applications for inert\footnote{The descriptor 
``inert'' is applied to the additional scalar doublet in the IDM 
in order to indicate that it does not couple to the SM fermions.   
We will therefore continue to refer to the fields this doublet
comprises as ``inert'' particles, even though these particles are not truly
inert in the sense that they have SM gauge interactions.} doublets exist
as well.  These range from explaining the lightness of neutrino 
masses via a one-loop radiative see-saw 
mechanism~\cite{IDMSeesaw}  to the loop-level induction of electroweak-symmetry 
breaking~\cite{BelgiansEWSBIDM} to engineering successful grand 
unification~\cite{IDMUnification}.  Furthermore, the model yields a
natural dark matter candidate in the form of the lightest 
inert particle (LIP), whose absolute stability is 
guaranteed by an unbroken $\mathbb{Z}_2$ symmetry.  Studies
of the relic abundance of LIP dark 
matter~\cite{BelgiansDarkIDM,ArizonansDarkIDM}, as well
as its prospects for indirect detection via neutrino~\cite{NeutrinoDarkIDM}, 
cosmic-ray positron and antiproton~\cite{BelgiansPositronIDM}, and
gamma-ray~\cite{SwedesGammaRayIDM,ArizonansGammaRayIDM} signatures,
and for direct detection~\cite{DirectDetectIDM} have also been performed.
  
Since the coupling structure of the fields of the additional 
scalar doublet in the IDM differs from that of typical two-Higgs doublet 
models (2HDM) in the manner discussed above, the collider phenomenology 
of the IDM also differs markedly from that of such 2HDM.  It is therefore
worthwhile to investigate the prospects for detecting the additional
fields of the IDM via their decay signatures at the Large Hadron 
Collider (LHC).  In this work, we focus on the dilepton (plus missing energy) 
channel, which turns out to be one 
of the most auspicious channels in terms of its discovery potential.  Some
preliminary, parton-level studies of this channel have been 
conducted~\cite{CaoMa} within one particular region of parameter space.  
Here, we present a more comprehensive, detector-level analysis in which
we investigate a variety of different benchmark regions motivated by dark-matter
studies, etc., and 
assess the prospects for observing a 
$\ell^+\ell^- + \displaystyle{\not}E_T$ signal at the LHC in each regime.
We note that the results of this analysis, although conducted 
in the context of the IDM, should also be applicable to other extensions of 
the SM with similarly-modified scalar sectors, as long as the extra scalars 
in those extensions have similar decay patterns to those in the IDM.

We begin in Sect.~\ref{sec:Model} with a recapitulation of the Inert
Doublet Model.  In Sect.~\ref{sec:Constraints}, we summarize 
the theoretical and experimental constraints to 
which the model is subject.   We outline a set of representative 
benchmark points which correspond to phenomenologically interesting
regions of the parameter space in which all of these constraints 
are satisfied.  In Sect.~\ref{sec:Cuts}, we  discuss
dilepton production in the IDM, as well as the SM backgrounds  
for $\ell^+\ell^- + \displaystyle{\not}E_T$ at the LHC, and 
we outline the event-selection criteria we use to differentiate
signal events from those produced by these backgrounds. 
We present our numerical results  in Sect.~\ref{sec:Results},
and in  Sect.~\ref{sec:Conclusion}, we conclude.


\section{Model Framework\label{sec:Model}}

The Inert Doublet Model is an extension of the SM 
in which the Higgs sector comprises two scalar ${\rm SU}(2)$ doublets.  The first 
of these doublets, here denoted $\phi_1$, resembles the SM Higgs doublet $H_{\rm SM}$.  
It bears full responsibility for electroweak-symmetry breaking and its neutral component acquires a vacuum expectation value (VEV) equal to that of the SM Higgs: 
$\langle\phi_1^0\rangle = v /\sqrt{2}= 174$~GeV.  It also couples to 
quarks and leptons in precisely the way $H_{\rm SM}$ does in the SM.
The second doublet, here denoted
$\phi_2$, does not contribute to EWSB ($\langle\phi_2\rangle = 0$), nor does it
have Yukawa interactions with the SM quarks and leptons.  This coupling 
structure is enforced by a $\mathbb{Z}_2$ symmetry under which
\begin{equation}
   \phi_1\rightarrow\phi_1,~~~~~\phi_2\rightarrow -\phi_2,
\end{equation}
and all the SM fields transform trivially.  This symmetry, sometimes called
matter parity, remains unbroken in the vacuum, since $\langle\phi_2\rangle$=0.
The most general $CP$-even scalar potential allowed by this $\mathbb{Z}_2$ symmetry   is
\begin{equation}
  V = \mu_1^2|\phi_1|^2 + \mu_2^2|\phi_2|^2 +
      \lambda_1|\phi_1|^4 + \lambda_2|\phi_2|^4 +
      \lambda_3|\phi_1|^2|\phi_2|^2 + \lambda_4 |\phi_1^\dagger\phi_2|^2 +
      \left[\frac{\lambda_5}{2}(\phi_1^\dagger\phi_2)^2 + \mathrm{h.c.}\right].
      \label{eq:V}
\end{equation}

After EWSB is triggered by the VEV of $\phi_1$, 
the scalar spectrum of the model comprises the usual SM Higgs 
$h$ (the neutral, $CP$-even degree of freedom in $\phi_1$), as
well as four additional fields corresponding to the four degrees of freedom in
$\phi_2$.  These include a pair of charged scalars $H^\pm$, 
a neutral, $CP$-even scalar $S$, 
and a neutral, $CP$-odd scalar $A$.  The masses of these scalars, given in terms
of the six free parameters\footnote{The seventh parameter $\mu_1^2$ appearing 
in Eq.~(\ref{eq:V}) is fixed by the constraint $v^2 = -\mu_1^2/\lambda_1$ from EWSB.} 
$\{\mu_2^2,\lambda_1,\lambda_2,\lambda_3,\lambda_4,\lambda_5\}$ in 
Eq.~(\ref{eq:V}), are
\begin{eqnarray}
  m_h^2 &=& 2\lambda_1 v^2, \\
  m_{H^{\pm}}^2 &=& \mu_2^2 + \lambda_3 v^2/2, \\
  m_S^2 &=& \mu_2^2 + (\lambda_3 + \lambda_4 + \lambda_5) v^2, \\
  m_A^2 &=& \mu_2^2 + (\lambda_3 + \lambda_4 - \lambda_5) v^2/2.
\end{eqnarray}
For our purposes, it will be useful to parametrize 
the model using the alternative parameter set 
$\{m_h,m_S,\delta_1,\delta_2,\lambda_2,\lambda_L\}$, where
$\delta_1 \equiv m_{H^\pm}-m_S$, $\delta_2 \equiv m_A-m_S$, and
$\lambda_L \equiv \lambda_3 + \lambda_4 + \lambda_5$.  This
parameterization is particularly useful in that it
characterizes the model in terms of physically significant 
quantities such as particle masses, mass splittings, and 
$\lambda_L$: the coefficient which controls the trilinear 
$hSS$ and quartic $hhSS$ couplings.            

The assumption of an unbroken $\mathbb{Z}_2$ matter parity
renders at least one of the degrees of freedom in $\phi_2$ 
absolutely stable. If this particle is electrically 
neutral ($S$ or $A$ in the IDM), 
it becomes a good weakly-interacting-massive-particle (WIMP) dark-matter candidate.  
For the remainder of this work, we will assume that the 
LIP is the $CP$-even scalar $S$, and hence $\delta_2>0$.   
The phenomenology of the alternative scenario, in which the $CP$-odd
scalar $A$ plays the role of the LIP, is very similar.

\section{Model Constraints\label{sec:Constraints}}

A variety of considerations, stemming both from theoretical consistency 
conditions and from experimental bounds, constrain the IDM.  Below, we briefly 
summarize these constraints, which were discussed in detail in 
Ref.~\cite{ArizonansDarkIDM}.

\begin{itemize}
\item{\bf Perturbativity:}
\begin{equation}
  \lambda_3^2 + (\lambda_3 + \lambda_4)^2+\lambda_5^2 < 12\lambda_1^2,
    ~~~~~~~~~~~~~~~~ \lambda_2 <1.
  \label{eq:Perturb1}
\end{equation}

\item{\bf Vacuum stability:}
\begin{eqnarray}
  \lambda_{1} > 0, ~~~~~~ ~~~~~ & ~~~~~ \lambda_2 > 0,\nonumber \\
  \lambda_3 > -2\sqrt{\lambda_1\lambda_2}, ~~~~~ & ~~~~~
  \lambda_3 + \lambda_4 - |\lambda_5| > -2\sqrt{\lambda_1\lambda_2}.
\end{eqnarray}  

\item{\bf Limits from direct collider searches:}

First of all, the excellent agreement between the experimentally-measured values for 
$\Gamma_W$ and $\Gamma_Z$ obtained from LEP and Tevatron 
data~\cite{WZWidthMeas} and the predictions of the SM requires that
\begin{eqnarray}
  2m_S +\delta_1 > M_W,~~~~~~&~~~~~~2m_S + \delta_1 + \delta_2 > M_W,\nonumber\\
  2m_S + \delta_2 > M_Z,~~~~~~&~~~~~~~2m_S + 2\delta_1>M_Z,
\end{eqnarray} 
in order that the decays $W^\pm\rightarrow S H^\pm, AH^\pm$  and $Z\rightarrow S A, H^+H^-$ 
are kinematically forbidden.  

Second of all, bounds on the invisible decays of 
the Higgs boson from LEP data~\cite{LEPInvisHWidth} also serve to constrain scenarios 
in which the Higgs is light and $m_h > 2m_S$.  In this work, however, we will be 
primarily concerned with cases in which $m_h>114$~GeV, for which 
the bounds from the searches on invisible Higgs decay does not apply.
 
Third and finally, one can consider limits arising from direct searches for
$H^\pm$, $A$, and $S$, both at LEP and at the 
Tevatron~\cite{HpLEPBound,HpTevatronBound}.  It should first be noted that 
the standard limits on additional charged and neutral Higgs scalars 
do not apply, because the standard search channels from which 
they are derived generally involve the couplings of such scalars to fermions, 
which are absent in the IDM.  On the other hand, bounds derived from the 
non-observation of $e^+e^-\rightarrow\chi_1^0\chi_2^0$~\cite{LEPSUSYNeutralino} and
$e^+e^-\rightarrow\chi^+_1\chi^-_1$~\cite{LEPSUSYChargino} decays in 
supersymmetric models can be used to constrain the IDM parameter space, since 
$e^+e^-\rightarrow SA$ and $e^+e^-\rightarrow H^+H^-$ in the IDM lead to similar signals.
A detailed analysis of the constraints on $e^+e^-\rightarrow SA$ in the IDM based 
on LEP~II searches for $e^+e^-\rightarrow\chi_1^0\chi_2^0$ 
was conducted in Ref.~\cite{SwedesIDMLEPII}, which showed that regions of 
parameter space with $m_S\lesssim 80$~GeV and $m_A\lesssim 100$~GeV for 
$\delta_2 \gtrsim 8$~GeV had been ruled out.  For 
$\delta_2 \lesssim 8$ GeV, however, only the LEP~I constraint $m_S+m_A > M_Z$ 
applies.  A rough bound of $m_{H^\pm}\gtrsim 70 - 90$~GeV~\cite{PierceThaler} 
can also be derived from the LEP $e^+e^-\rightarrow\chi^+_1\chi^-_1$ limit by making
the necessary modifications to account for the difference in cross-section
between fermion-pair and scalar-pair production.  Taking these considerations
into account, we will henceforth restrict our attention to models for 
which $m_{H^\pm}>80$~GeV.

\item{\bf Electroweak precision constraints:}

Electroweak precision measurements set limits on contributions from the additional Higgs doublet to the oblique $S$ and $T$ parameters~\cite{STU}.  We consider
a given parameter choice to be consistent with electroweak precision 
constraints as long as the overall values of $S$ and $T$ it yields, once
all contributions are taken into account, lie within the 68\%~C.L. ellipse 
determined by the LEP Electroweak Working Group~\cite{STUContLEP}.
For a light SM Higgs, with $m_h\lesssim 200$ GeV, the constraint is weak as long as 
$\delta_1$ and $\delta_2$ are of roughly the same order.  For a heavy SM Higgs, 
a large splitting between $H^\pm$ and $S$ is preferred, and $\delta_1 > \delta_2$.

\item{\bf Dark matter relic density:}

One of the attractive aspects of the Inert Doublet Model is
that it can provide a viable WIMP dark matter candidate in the
form of a stable, neutral LIP.   
The model is therefore constrained by experimental limits on the relic 
density of dark matter in the universe.   
In what follows, we will assume that the LIP relic density represents 
the dominant component of $\Omega_{\mathrm{DM}}h^2$
and falls within the $3\sigma$ range of the dark-matter 
density of the universe as measured by WMAP~\cite{WMAP2003}\footnote{In
the event that additional sources contribute to $\Omega_{\mathrm{DM}}h^2$,
only the upper bound applies.}:  $0.085<\Omega_{\mathrm{DM}}h^2<0.139$.
A detailed examination of the relic density of a $CP$-even scalar LIP
in the IDM was conducted in~\cite{ArizonansDarkIDM}.  It was 
found that the correct dark-matter relic density could be realized in 
several distinct regions of parameter space in which   all the 
aforementioned theoretical and experimental constraints were also satisfied.  
For a light SM Higgs with $m_h\sim 120$~GeV, two scenarios are possible. 
The first of these involves a light LIP
with $m_S \sim 40 - 80$~GeV  
and mass splittings $\delta_{1}$ and $\delta_{2}$ which
are sizeable, but of the same order.  
The second involves a heavier dark matter particle with 
$m_S\gtrsim 400$ GeV and relatively small mass splittings.
For a heavy SM Higgs with $m_h\gtrsim 400$ GeV, the regions which
the constraints leave open are those 
in which $m_S \sim 80$~GeV and $\delta_1 > \delta_2$,
with both $\delta_1$ and $\delta_2$ relatively large, or
$m_S \sim 50 - 75$~GeV, $\delta_2 \lesssim 8$ GeV with a large $\delta_1$.
 
\end{itemize}

\begin{table}
\begin{center}
\begin{tabular}{|c|c|c|c|c|c|}\hline
~~Benchmark~~ & ~$m_h$ (GeV)~& ~$m_S$ (GeV)~ & ~$\delta_1$ (GeV)~ & 
~$\delta_2$ (GeV)~ & ~~~~$\lambda_L$~~~~  \\\hline
LH1& 150 &  40 & 100 & 100 & $-0.275$\\
LH2& 120 &  40 &  70 &  70 & $-0.15$ \\
LH3& 120 &  82 &  50 &  50 & $-0.20$ \\
LH4& 120 &  73 &  10 &  50 & $0.0$   \\
LH5& 120 &  79 &  50 &  10 & $-0.18$ \\ \hline
HH1& 500 &  76 & 250 & 100 & $0.0$   \\
HH2& 500 &  76 & 225 &  70 & $0.0$   \\
HH3& 500 &  76 & 200 &  30 & $0.0$   \\\hline
\end{tabular}
\caption{A list of benchmark points used in our analysis, defined in terms of the model parameters $\{m_h,m_S,\delta_1,\delta_2,\lambda_L\}$.  Dark matter relic density and collider phenomenology of the IDM depend little on $\lambda_2$, which is set to 0.1  for all benchmark points.  The points LH1 $-$ LH5
involve a light ($120\mbox{ GeV}\leq m_h \leq 150\mbox{ GeV}$) Higgs boson, while the 
points HH1 $-$ HH3 involve a heavy ($m_h=500\mbox{ GeV}$) Higgs.}
\label{tab:BMs}
\end{center}
\end{table}

In Table~\ref{tab:BMs}, we define a set of 
benchmark points, each designed to represent a particular region of the 
remaining, ``habitable'' parameter space, with an eye toward its collider
phenomenology.  
We emphasize that each benchmark point in Table~\ref{tab:BMs} is consistent
with all of the applicable constraints detailed above, and that each
yields an LIP relic density that falls within the WMAP $3\sigma$ range for
$\Omega_{\mathrm{DM}}h^2$.

The first regime of interest involves a light SM Higgs with $m_h<200$ GeV. 
For such Higgs masses, as discussed above, the electroweak precision constraints are not 
terribly stringent, and a wide variety of possible particle spectra are permissible.  
We have included five different benchmark points in our analysis which correspond
to this regime (labeled LH1$ - $LH5 for ``light Higgs''), the properties
of which are listed in Table~\ref{tab:BMs}.  These points are representative of
the set of possible scenarios which differ qualitatively  from the perspective 
of a dilepton-channel analysis at the LHC.  
The points LH1 and LH2 both represent cases in which
the LIP is light ($\sim$ 40 GeV) and $\delta_1$ and $\delta_2$ are large 
and of the same order.  However, for the point LH1, $\delta_{1}>M_W$ and 
$\delta_{2}>M_Z$, meaning that both $H^\pm$ and $A$ 
can decay on shell (to $SW^\pm$ and $SZ$, respectively), whereas for LH2, 
$\delta_{1}<M_W$ and $\delta_{2}<M_Z$, so only three-body decay are 
kinematically accessible.  
A slightly larger Higgs mass $m_h=150$ GeV is mandated in LH1 by 
perturbativity constraints.  However, the collider phenomenology of $S$, 
$A$ and, $H^\pm$ --- at least as far as the dilepton channel is concerned --- does 
not depend significantly on the value of $m_h$, as will soon be made apparent.  

Points LH3$ - $LH5 all correspond to situations in which $m_S\sim 80$~GeV,
but each represents a different relationship between $\delta_1$ 
and $\delta_2$.  The point LH3
represents a situation similar to that embodied by LH2, in which $\delta_1$ 
and $\delta_2$ are on the same order and on-shell decays to $SA$ and $SW^\pm$ 
are inaccessible.  Larger values of $\delta_{1,2} > M_{W,Z}$ are disfavored by 
the aforementioned battery of constraints.   
The point LH4 represent the case of intermediate $\delta_2$ 
and small $\delta_1$, while the point LH5 represents the opposite situation, in 
which $\delta_1$ is of intermediate size and $\delta_2$ is small.  It is also 
possible to realize a situation similar to that of LH4, but with $\delta_2>M_{Z}$
and hence on-shell $A$ decay.  The dilepton-channel analysis in this case 
would be similar to that in LH1 and HH1.  Another 
possibility would be a point similar to LH5, but
with $\delta_1>M_W$, so that on-shell $H^\pm$ decays would be allowed.  
However, as will be explained in more detail below, the dilepton-signal contribution 
from $H^+H^-$ decay is hard to disentangle from the SM $W^+W^-$ background.
Consequently, varying $\delta_1$ has little effect on the observability of the 
dilepton signal via $SA$ associated production, by far the most useful production
process for discovery at the LHC.

The second regime of interest involves a heavy SM Higgs with $m_h\gtrsim 400$~GeV. 
A large splitting between $H^\pm$ and $S$ is required to satisfy the constraints 
from electroweak precision measurements in this case.  Broadly speaking, these
constraints, taken in tandem with relic-abundance considerations, prefer 
$\delta_1$ to be quite large (and generally much larger than $\delta_2$) 
and the LIP mass to lie within the range 
$m_S\approx 70 - 80$~GeV~\cite{ArizonansDarkIDM}.  This parameter-space 
regime is represented by the benchmark points HH1$ - $HH3 (where ``HH'' stands
for ``Heavy Higgs'') in Table~\ref{tab:BMs}.   The point 
labeled HH1 represents the case in which $\delta_2>M_Z$ and 
$A$ decays proceed via an on-shell $Z$ intermediary, while the point 
HH2 represents the case in which $\delta_2<M_Z$, and the decay $A\rightarrow SZ$
is kinematically inaccessible.  
HH3 is similar to HH2, but has a small $\delta_2=$ 30 GeV.
Since precision constraints generally
dictate that $m_H^{\pm}>m_A>m_S$ if $S$ is to be the LIP, these three cases
encapsulate the only qualitatively different possibilities in this regime from the
perspective of dilepton-channel analysis.  It is worth noting that another   
region of parameter space does exist in which all the aforementioned constraints are 
satisfied: one in which $50~\mbox{GeV}\lesssim m_S \lesssim 75$~GeV 
and the mass splitting $\delta_2$ is very small ($\delta_2 \lesssim 8$~GeV). 
However, a dilepton signal tends to be exceedingly difficult to observe in 
scenarios of this sort, due to the softness of the jets and leptons in the final 
states.  For this reason, we do not include a representative benchmark point for 
this region in the present study.  
 
For the other allowed region of parameter space --- 
that in which $m_S\gtrsim 400$~GeV and the mass splittings $\delta_{1}$ and 
$\delta_{2}$ are relatively small --- no benchmark points have been included in 
this study.  This is because a scenario of this sort does not yield a 
detectable signal in the dilepton channel.  One reason for this is that 
the pair-production cross-sections for the inert scalars are highly 
suppressed  due to their heavy masses.  Another is that the jets and 
leptons produced during $H^\pm$ and $A$ decays will be quite soft, owing to the 
small size of the mass splittings.  Therefore, although it remains a
phenomenologically viable scenario, we will not discuss this possibility further 
in the present work.      


\section{Signals, Backgrounds, and Event Selection\label{sec:Cuts}}

Let us now turn to examine the signal and background processes relevant
to an analysis of the dilepton channel in the IDM at the LHC. 
The inert scalars $H^{\pm}$, $A$, and $S$
can be pair-produced directly at the LHC by Drell-Yan processes involving
virtual photons and $W^{\pm},Z$ bosons:
\begin{eqnarray}
  q\bar{q}\rightarrow Z\rightarrow AS\nonumber, ~~~~~~~~ & ~~~~~
  q\bar{q}\rightarrow Z/\gamma^\ast\rightarrow H^+H^-, \nonumber \\ 
  q\bar{q}'\rightarrow W^\pm\rightarrow AH^\pm, ~~~~~ & ~~~~~
  q\bar{q}'\rightarrow W^\pm\rightarrow SH^\pm.
  \label{eq:SigProcs}
\end{eqnarray}
In Table~\ref{tab:BMXSecs}, 
we listed the production cross-sections for $SA$, $H^+H^-$, $SH^\pm$, and $A H^\pm$
at the LHC for the various benchmark points defined in Table~\ref{tab:BMs}.
Once so produced, the unstable $H^\pm$ and $A$ bosons further decay to 
lighter states plus $W^{(*)}$ or $Z^{(*)}$.  Depending on how $H^\pm$ and 
$A$ decay, a number of final states are possible.  Each of these states, as 
required by matter parity, includes precisely two LIPs, as well as a number of
jets, charged leptons, and neutrinos.

\begin{table}
\begin{center}
\begin{tabular}{|c|cccc|}\hline
~~Benchmark~~ &~$\sigma_{SA}$ ~&~$\sigma_{H^+H^-}$ ~ &
~$\sigma_{SH^\pm}$ ~&~$\sigma_{AH^\pm}$ ~ \\
&(fb)&(fb)& (fb)&(fb)\\          
\hline
LH1 & 289.2 &  69.8 &  503.3 & 125.2 \\
LH2 & 628.8 & 163.6 & 1055.1 & 299.0 \\
LH3 & 179.9 &  86.0 &  319.0 & 154.9 \\
LH4 & 248.9 & 440.2 & 1050.3 & 370.1 \\
LH5 & 465.5 &  93.3 &  352.9 & 302.3 \\\hline
HH1 &  91.8 &   2.9 &   25.4 &  13.5 \\
HH2 & 152.0 &   4.0 &   33.0 &  20.5 \\
HH3 & 336.7 &   5.6 &   43.7 &  35.2 \\\hline
\end{tabular}
\caption{Leading-order cross-sections for the associated production of 
$SA$, $H^+H^-$, $SH^\pm$, and $AH^\pm$ at the 
LHC, with center-of-mass energy $\sqrt{s}=14$~TeV, 
for the various benchmark points 
defined in Table~\ref{tab:BMs}.
\label{tab:BMXSecs}}
\end{center}
\end{table}

The presence of sizeable QCD backgrounds for final states involving one
or more jets renders such states difficult to use for discovery; final
states involving charged leptons alone, on the other hand, have far
smaller SM backgrounds and hence are far more auspicious in terms of their
LHC discovery potential.  
A single lepton plus missing $E_T$ signal would be difficult to resolve, 
due to the huge SM $W$ background, but a variety of multi-lepton signatures 
initiated by the electroweak processes enumerated above may be observable 
at the LHC.  The trilepton + $\displaystyle{\not}E_T$ channel, for example, which 
is of crucial importance for supersymmetry searches~\cite{trilepton}, can
potentially also be important in searching for an additional, inert 
scalar doublet.  In this work, however, we will focus on  dilepton channel, which seems to offer the brightest prospects for discovery.

\begin{figure}[bht]
\begin{center}
\resizebox{2.0in}{!}{\includegraphics{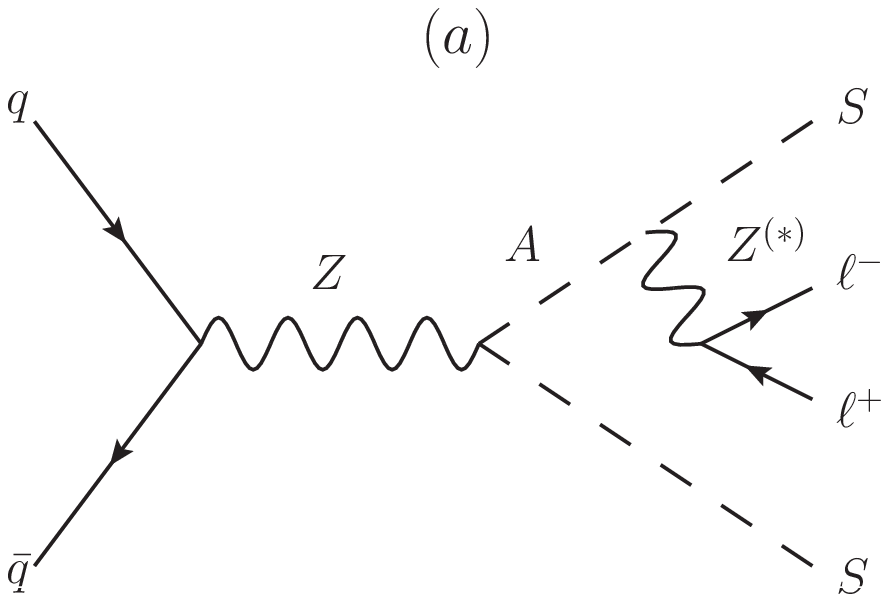}}~~~~~~~~~
\resizebox{2.0in}{!}{\includegraphics{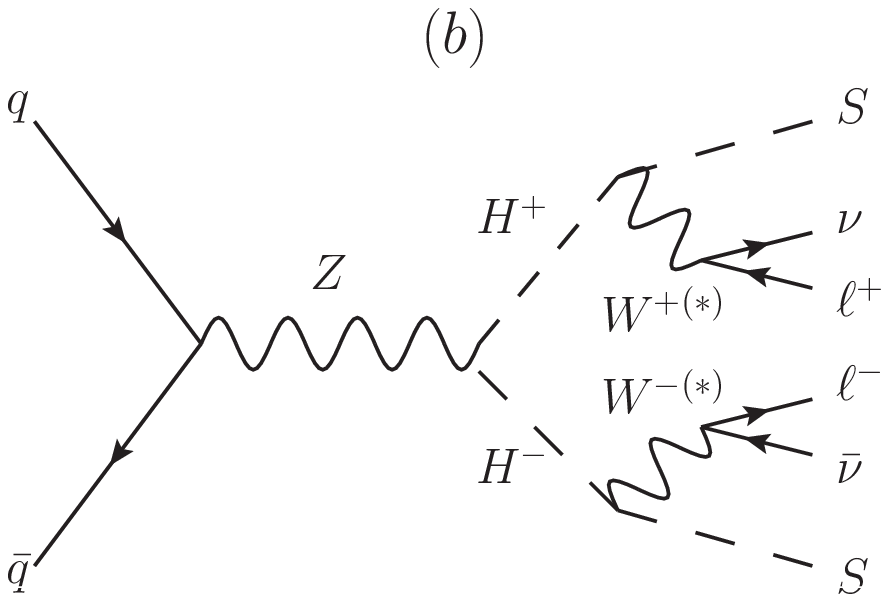}}\\
\vspace{.5cm}
\resizebox{2.0in}{!}{\includegraphics{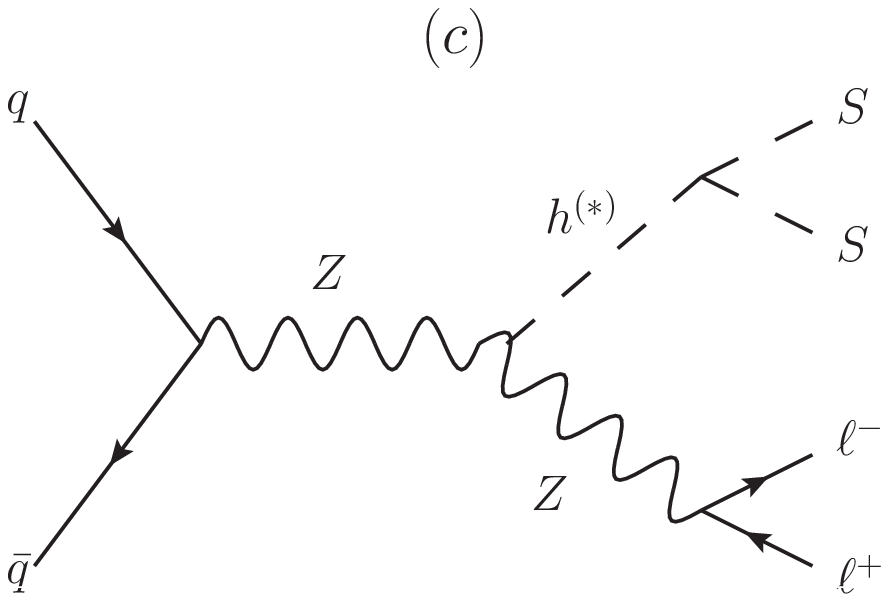}}~~~~~~~~~
\resizebox{2.0in}{!}{\includegraphics{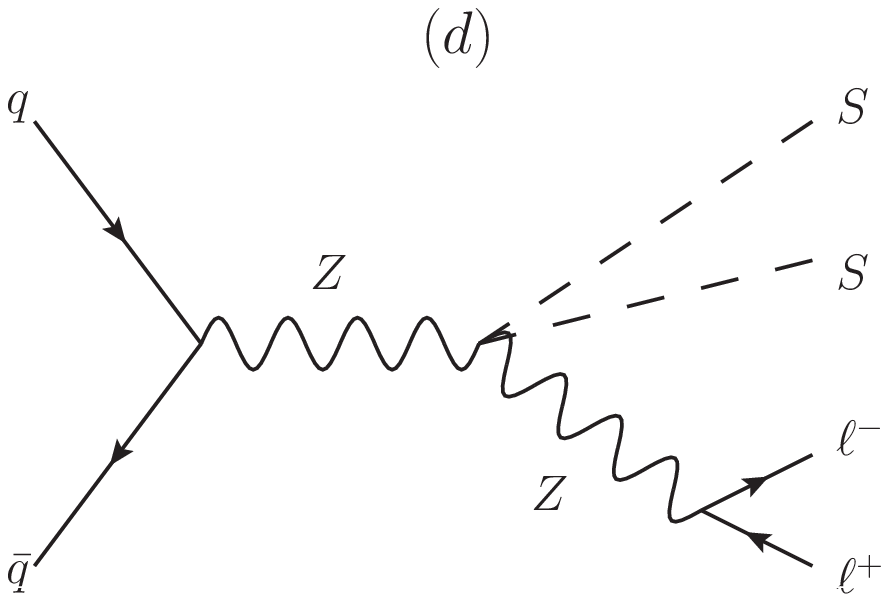}}
\caption{Diagrams corresponding to the  contributions to the
$pp\rightarrow \ell^+\ell^-\displaystyle{\not}E_T$   in the IDM
discussed in the text.  
\label{fig:feyn}}
\end{center}
\end{figure}

The dominant signal contribution to $\ell^+\ell^- + \displaystyle{\not}E_T$ 
in the IDM, where $\ell=\{e,\mu\}$ ,
results from either $pp\rightarrow SA$ with $A\rightarrow S\ell^+\ell^-$, 
or $pp\rightarrow H^+H^-$, with $H^{\pm}\rightarrow S\ell^\pm\nu$, depending on
the choice of parameters.  These processes are depicted diagrammatically in panels 
(a) and (b) of Fig.~\ref{fig:feyn}, respectively.  Other processes that result in
$\ell^+\ell^-+\displaystyle{\not}E_T$ final states, e.g.\ $pp\rightarrow H^+H^-$ 
with $H^+\rightarrow S\ell^+\nu$ and 
$H^-\rightarrow A\ell^-\bar{\nu}\rightarrow S\ell^-\bar{\nu}\nu\bar{\nu}$, 
generally contribute only a small amount to the signal cross-section 
and can therefore be safely ignored.  Another contribution comes from processes in 
which a leptonically-decaying pseudoscalar $A$ is produced
in association with some other particle or particles which decay to
jets or charged leptons too soft to register in the detector.  In general, the event rates 
for such processes (the most important of which is 
$pp\rightarrow H^\pm A\rightarrow \ell^+\ell^-jj+\displaystyle{\not}E_T$) 
are also small compared to that for 
$pp\rightarrow SA\rightarrow \ell^+\ell^-+\displaystyle{\not}E_T$.  However,     
if $\delta_1$ is small (as it is in benckmark LH4), a substantial fraction 
of the jets and charged leptons from $H^\pm$ decays will be sufficiently soft 
that such processes do yield a considerable contribution and therefore need
to be accounted for in the analysis.       

In addition to the pair-production processes discussed above,
the electroweak Higgs-associated-production process 
\begin{equation}
  q\bar{q}\rightarrow h Z 
\end{equation} 
can also result in a $\ell^+\ell^-+\displaystyle{\not}E_T$ final
state in the manner illustrated in panel (c) of Fig.~\ref{fig:feyn}, as long as 
the decay $h\rightarrow SS$ is permitted.  The dilepton-channel 
contribution from this process is significant only in cases in which 
$\lambda_L$ is nonzero and $m_h>2 m_S$ --- the conditions under which 
$h$ can decay on-shell to a pair of LIPs.  Of the eight 
benchmark scenarios defined in Table.~\ref{tab:BMs}, these conditions 
are satisfied only in scenarios LH1 and LH2, for which 
$\sigma_{hZ}\times {\rm Br}(h\rightarrow SS)=343.12$~fb 
and 706.65~fb, respectively. 
As these rates are roughly on the same order
as those for $pp\rightarrow SA$ production, it will be necessary to take 
this contribution into account in the ensuing analysis.     

A further contribution to the $\ell^+\ell^-\displaystyle{\not}E_T$ 
production cross-section in the IDM results from $pp\rightarrow SSZ$ 
production via the four-point $SSZZ$ interaction shown in 
panel (d) of Fig.~\ref{fig:feyn}.  
However, this contribution is quite small in comparison with that from $SA$ 
pair production, as the former is a three-body process while the latter 
is only two-body.  Interference effects between the
diagrams depicted in panels (a), (c), and (d) of Fig.~\ref{fig:feyn} are 
consequently tiny as well, and can be safely neglected.
 
In what follows, we will focus on 
$pp\rightarrow SA\rightarrow\ell^+\ell^-SS$ as our signal process, and treat
$pp\rightarrow H^+H^-\rightarrow\ell^+\ell^-\nu\bar{\nu} SS$ as part of the background.  
The reason for this is twofold.  First, since the  constraints in
Sect.~\ref{sec:Constraints} (and especially those from WMAP and 
electroweak precision data) typically prefer situations in 
which $2m_{H^+}\gtrsim m_A + m_S$,
production cross-sections for $pp\rightarrow H^+H^-$ tend to be lower 
than those for $pp\rightarrow SA$.  Second, a 
$pp\rightarrow H^+H^-\rightarrow\ell^+\ell^-+\displaystyle{\not}E_T$ 
signal turns out to be far more difficult to distinguish from the dominant 
SM backgrounds (discussed in detail below) on the basis of event topology. 
We will also treat
$pp\rightarrow h Z\rightarrow SS \ell^+\ell^-$ and $pp\rightarrow SSZ\rightarrow SS \ell^+\ell^-$ as a contribution to 
the background,  because the event topologies generally differ
from those associated with $pp\rightarrow SA\rightarrow\ell^+\ell^-SS$.  

The SM backgrounds relevant for the $\ell^+\ell^- + \displaystyle{\not}E_T$
channel are well-known from studies of the supersymmetric process
$pp\rightarrow\chi_1^0\chi_2^0\rightarrow\ell^+\ell^- +\displaystyle{\not}E_T$, 
where $\chi_1^0$ and $\chi_2^0$ are the lightest and next-to-lightest
neutralinos.   
These include irreducible backgrounds from $WW$ and $ZZ/\gamma^\ast$ 
production (with the contribution from off-shell photons~\cite{ZGammaDilepBG}
properly taken into account), as well as reducible 
backgrounds from $t\bar{t}$, 
$WZ/\gamma^\ast$, $Wt$, and $Zb\bar{b}$ processes; $WW+n\mbox{ jets}$ and 
$ZZ+n\mbox{ jets}$; and Drell-Yan production of
$\tau^+\tau^-$ pairs.  

In the present analysis, events were generated at parton-level, both for the
signal process and for the backgrounds discussed above, in MadGraph~\cite{MadGraph} 
and then passed through PYTHIA~\cite{PYTHIA} for parton showering and hadronization.  
Events were then passed through PGS4~\cite{pgs} to simulate the effects 
of a realistic detector.  Subsequent to event generation,
in order to distinguish signal events from those associated with these  
backgrounds and to account for the performance thresholds of the 
LHC detectors, we impose three sets of cuts in our analysis.    
The first such set, henceforth referred to as our 
Level~I cuts, is designed to mimic a realistic detector
acceptance:  
\begin{itemize}
\item Exactly two electrons or muons with opposite charge.
\item $p_T^{\ell} \geq 15$~GeV and $|\eta_{\ell}|\leq 2.5$ for each
   of these charged leptons.
\item For lepton isolation, we require $\Delta R_{\ell\ell} \geq 0.4$ 
  for the charged-lepton pair, 
  and $\Delta R_{\ell j} \geq 0.4$ for each combination of one jet 
  and one charged lepton.  
\end{itemize}
It should be noted that for $\ell=\{e,\mu\}$,  the above lepton $p_T^{\ell}$ cut 
is sufficient to meet the Level~I triggering requirements for both 
ATLAS~\cite{ATLASTDR} and CMS~\cite{CMSTDR}. 

The subsequent two sets of selection criteria we impose are designed to 
discriminate efficiently between signal and background events.  Our Level~II 
cuts are aimed at suppressing reducible backgrounds from processes
such as $t\bar{t}$, $WZ/\gamma^\ast$, $Wt$, and $Zb\bar{b}$, which tend to involve either 
hard jets, little missing transverse energy, or both.   
We impose a veto on all events manifesting 
high-$p_T$ jet activity within the central region of the detector, as well as a 
minimum missing transverse energy cut: 
\begin{itemize}
\item No jets with $p_T^j>20$~GeV and pseudorapidity within the range
  $|\eta_j|<3.0$.
\item $\displaystyle{\not}E_T > 30$~GeV.
\end{itemize}   
The efficacy of this latter missing $E_T$ cut should not be overemphasized: while
each signal event necessarily includes a pair of LIPs, 
these particles tend to be produced back-to-back.  As a result, 
their contributions to the overall $\displaystyle{\not}E_T$ tend
to cancel each other out, to the end that the $\displaystyle{\not}E_T$ 
distributions for signal events tend not to differ radically from those for
processes like $ZZ/\gamma^\ast$, $WW$, and $t\bar{t}$ which 
involve energetic neutrinos.  Nevertheless, this $\displaystyle{\not}E_T$ 
cut is highly efficient in eliminating background contributions from 
$Zb\bar{b}$ and $WZ/\gamma^\ast$ events (with the $W$ decaying hadronically) with 
jets soft enough so as to survive the central-jet veto.       

After imposing Level I and II cuts, 
contributions from $Zb\bar{b}$ and Drell-Yan 
production of leptonically-decaying $\tau^+\tau^-$ pairs, are effectively 
eliminated.  The dominant remaining 
backgrounds are the irreducible ones from $WW$ and 
$ZZ/\gamma^\ast$ events, as well as residual $t\bar{t}$, $WZ/\gamma^\ast$  
and $Wt$ events which survive the Level~II cuts.  
In Table~\ref{tab:BMXSecscutI_II}, we list the signal cross-sections
for $pp\rightarrow SA\rightarrow \ell^+\ell^-  \displaystyle{\not} E_T$ 
at the LHC for each of the benchmark points presented in 
Table~\ref{tab:BMs}, after the application of the Level~I and Level~II cuts
discussed in the previous section.  
We also show the effect that these 
cuts have on the cross-sections for those background processes, both
reducible and irreducible, which remain at non-negligible levels after
the Level~II cuts have been applied: $WW$, $ZZ/\gamma^\ast$, $t\bar{t}$, 
$WZ/\gamma^\ast$, and $Wt$.  
Results for $pp\rightarrow H^+H^-\rightarrow \ell^+\ell^- \displaystyle{\not} E_T$ 
and $pp\rightarrow h^{(*)}Z \rightarrow SS \ell^+\ell^-$, also
treated as background processes in this analysis, are shown here as well.
It is evident from the data presented in Table~\ref{tab:BMXSecscutI_II} that the application of
the Level~I+II cuts results in a substantial reduction of the reducible backgrounds 
from $t\bar{t}$, $WZ/\gamma^\ast$, and $Wt$.  However, as efficient as these cuts are, 
the rates for these background processes (and especially from $t\bar{t}$) are 
large enough that a substantial number of events still survives them.  Consequently,
these reducible backgrounds cannot be neglected in the final analysis. 

\begin{table}
\begin{center}
\begin{tabular}{|c|ccc|ccc||c|c|c|}\hline
&\multicolumn{3}{c|}{Level~I Cuts} &\multicolumn{3}{c||}{Level~I+II Cuts} &
\multirow{3}{*}{\parbox{2.3cm}{SM \\ Backgrounds}}  & Level~I Cuts &Level~I+II Cuts\\  \cline{2-7}  \cline{9-10}
Benchmark &$\sigma_{SA}$ &$\sigma_{H^+H^-}$ & $\sigma_{hZ}$ &
$\sigma_{SA}$ &$\sigma_{H^+H^-}$ &$\sigma_{hZ}$ & & $\sigma_{BG}$&$\sigma_{BG}$  \\
&(fb)&(fb)&(fb)& (fb)&(fb)&(fb)& & (fb) &(fb)\\          
\hline
LH1 &~ 9.61 & 0.82    & 2.90 ~&~6.03     &0.46 &1.79~~&$WW$            & 621.44 &316.97\\
LH2 &~10.28 & 1.06    & 5.75 ~&~6.53     &0.51 &3.47~~&$ZZ/\gamma^\ast$& 132.09 & 76.46\\
LH3 &~ 2.32 & 0.34    & 0.01 ~&~1.47     &0.13 &0.01~~&$t\bar{t}$      &4531.51 & 58.87\\
LH4 &~ 3.84 & 0.19    & 0    ~&~2.07     &0.02 &0   ~~&$WZ/\gamma^\ast$& 113.97 & 51.85\\
LH5 &~ 0.38 & $\sim 0$& 0.01 ~&~$\sim 0$ &0.14 &0.01~~&$Wt$            & 709.14 & 52.11\\ 
\hline
HH1 &~ 3.23 & 0.02    & 0    ~&~1.97     &0.01 &0   ~~&
\multirow{3}{*}{\parbox{2.3cm}{Total SM \\ Background}} &
\multirow{3}{*}{6108.15} & \multirow{3}{*}{556.26} \\
HH2 &~ 3.01 & 0.03    & 0    ~&~1.81     &0.01 &0   ~~&&& \\
HH3 &~ 1.69 & 0.02    & 0    ~&~1.09     &0.01 &0   ~~&&& \\ \hline
\end{tabular}
\caption{Leading-order cross-sections for the signal processes 
$pp\rightarrow SA\rightarrow \ell^+\ell^-  \displaystyle{\not} E_T$ at the LHC
with $\sqrt{s}=14$~TeV 
after Level I and II cuts for each of the benchmark points presented in 
Table~\ref{tab:BMs}.  Also shown are the backgrounds 
$pp\rightarrow H^+H^-\rightarrow \ell^+\ell^-  \displaystyle{\not} E_T$,  
$pp\rightarrow h^{(*)}Z \rightarrow \ell^+\ell^-  \displaystyle{\not} E_T$,
$WW$, $ZZ/\gamma^*$, $t\bar{t}$, $WZ/\gamma^\ast$, $Wt$ after Level I+II cuts, 
as well as a total background cross-section.\label{tab:BMXSecscutI_II}}
\end{center}
\end{table}

In order to differentiate the $pp\rightarrow SA$ signal 
from these remaining backgrounds, it is necessary to impose a third level 
of event-selection criteria based largely on event topology, whose
thresholds can be adjusted to optimize significance of discovery 
in any given benchmark scenario.  
For the benchmark points included in our survey, the optimal pattern of
Level~III cuts generally falls into one of two categories, depending primarily on 
whether or not the decay of $A \rightarrow S Z^{(*)}$ occurs on shell.
  
\begin{figure}[bht]
\begin{center}
\resizebox{3.15in}{!}{\includegraphics*[83,220][510,562]{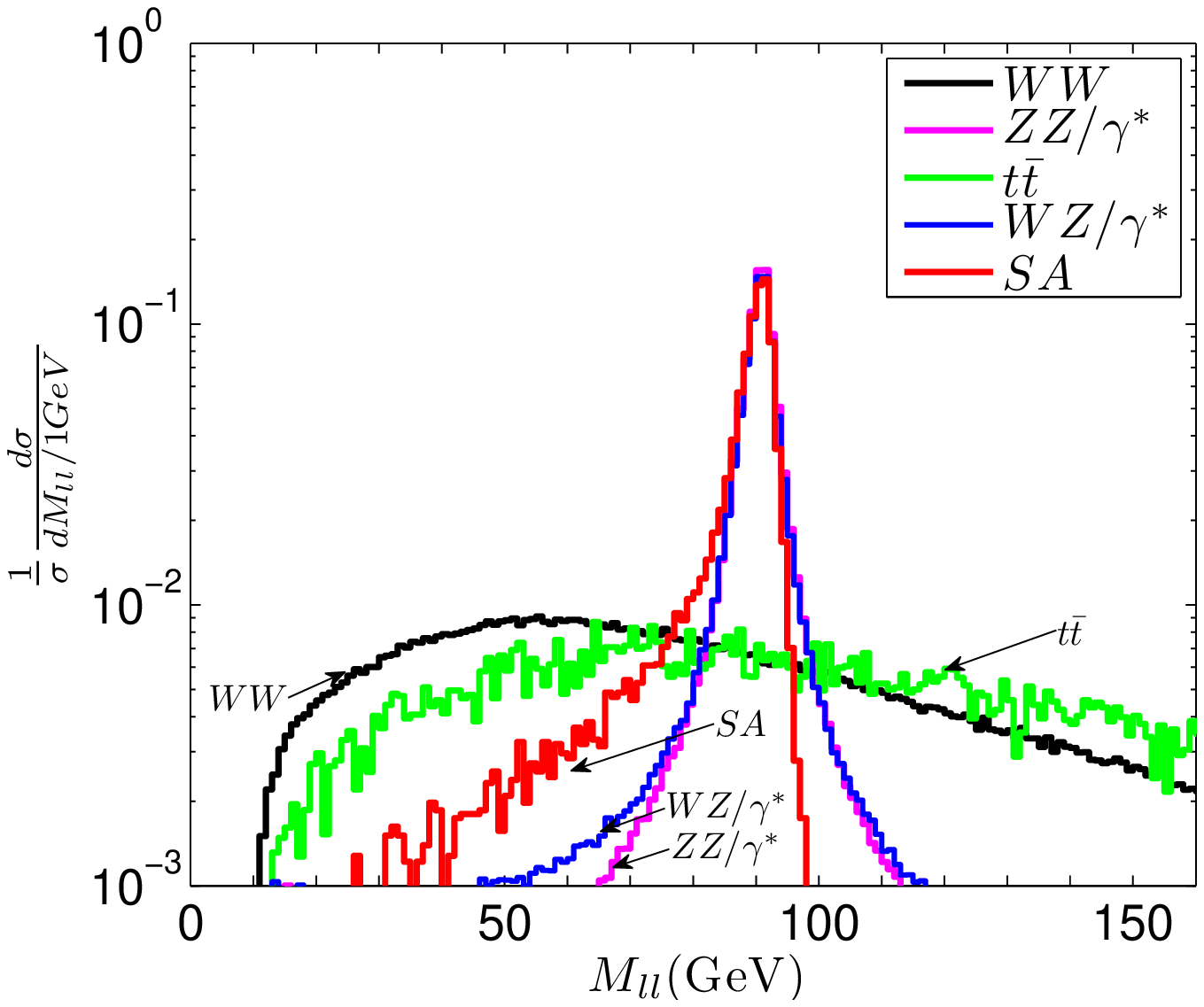}}
\resizebox{3.15in}{!}{\includegraphics*[83,220][510,562]{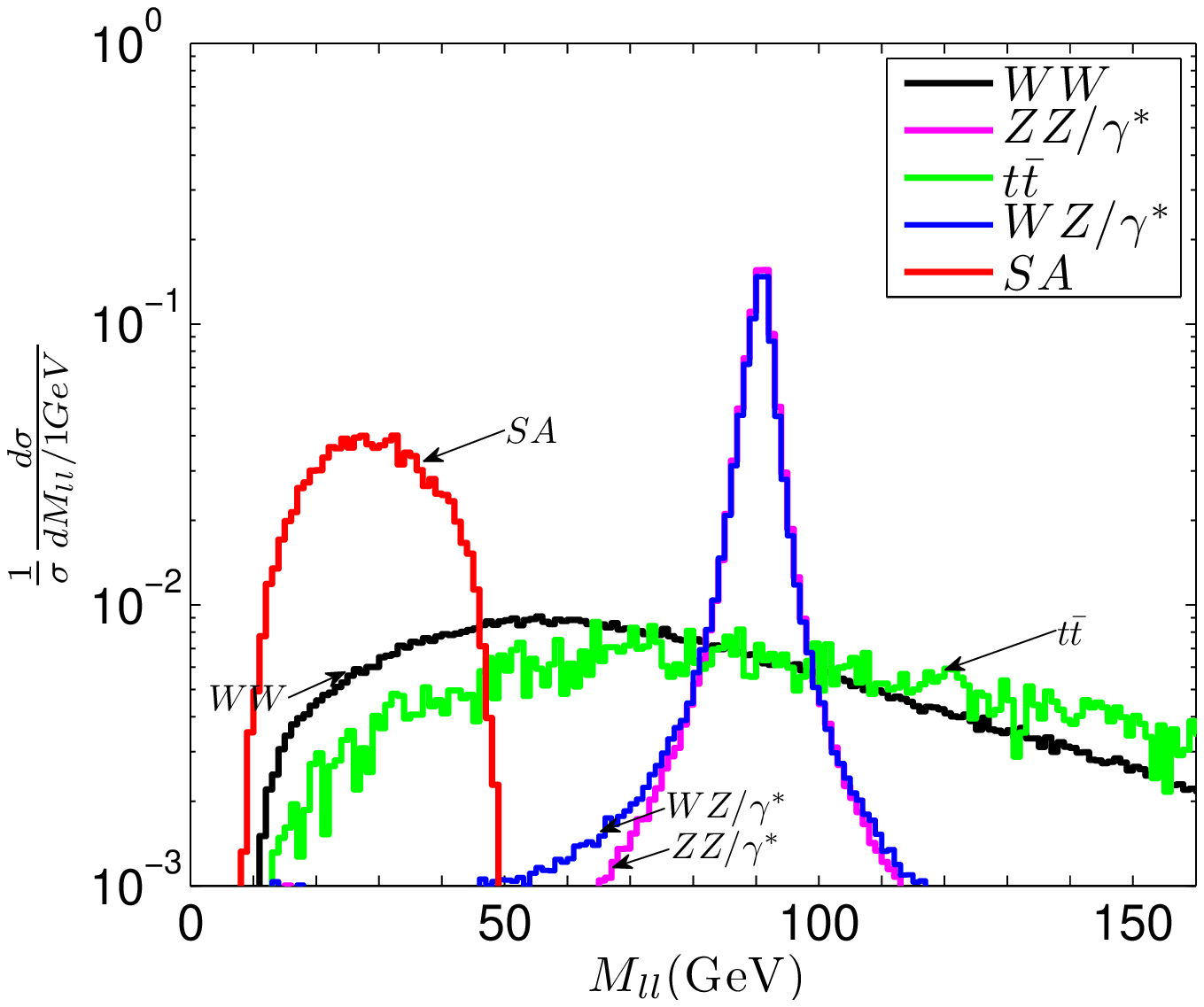}}
\caption{Dilepton-invariant-mass distributions for the benchmark points LH1 
(left panel) and LH3 (right panel) both for the signal process and for the 
most relevant SM backgrounds.}
\label{fig:Mll}
\end{center}
\end{figure}
 
In all of our benchmark scenarios in 
which $\delta_2>M_Z$, the $CP$-odd scalar $A$ decays 
essentially 100\% of the time to an LIP and an on-shell $Z$; thus the 
distribution of the invariant mass $M_{\ell\ell}$ of the charged-lepton
pair in such scenarios tends to peak sharply around $M_Z$.  This is the 
case for points LH1 and HH1, the $M_{\ell\ell}$ distribution
for the former of which is shown in the left panel of Fig.~\ref{fig:Mll}.
It is therefore advantageous to select events on the basis of
whether or not $M_{\ell\ell}$ falls within a window
\begin{itemize}
\item $M_{\ell\ell}^{\mathrm{min}}\leq M_{\ell\ell}
  \leq M_{\ell\ell}^{\mathrm{max}}$,
\end{itemize}  
where the parameters $M_{\ell\ell}^{\mathrm{min}}$ 
and $M_{\ell\ell}^{\mathrm{max}}$
are to be adjusted to optimize the statistical significance of
discovery for each benchmark point.  In cases of this sort, the
best results are generally obtained by imposing a window
around $20$~GeV wide, centered near $M_Z$.  
Such a cut efficiently
reduces the $W^+W^-$, $Z\gamma^\ast$, $W\gamma^\ast$, $t\bar{t}$ and $Wt$ backgrounds, 
leaving the $ZZ$ and $WZ$ backgrounds (which are little affected by such a cut) as
the dominant ones.  

In cases where $\delta_2<M_Z$ (LH2$ - $LH5, HH2$ - $HH3),
the two body decay $A\rightarrow SZ$ is kinematically inaccessible.
Likewise, the decay channel $A\rightarrow H^\pm W^\mp$ is not open
unless $\delta_2>\delta_1+M_W$ --- a condition which is difficult to
realize, given the constraints on the model, and which is not 
satisfied for any of the benchmark points in our study.  When these
decays are unavailable, the dominant leptonic decay channel for the $A$
involves the three-body process $A\rightarrow S\ell^+\ell^-$, which 
proceeds through an off-shell $Z$.  As a result, the dilepton invariant 
mass distribution is
peaked well below $M_Z$, around the value of $\delta_2$, as shown in the right panel 
of Fig.~\ref{fig:Mll} for benchmark point LH3.
In cases of this sort, imposing an upper limit 
$M_{\ell\ell}^{\mathrm{max}} \sim \delta_2$
on the dilepton invariant mass can assist in improving the 
signal-to-background ratio.  A cut of this sort can effectively suppress 
the $ZZ$ and $WZ$ backgrounds, the $M_{\ell\ell}$ distributions for which
are peaked sharply around $M_Z$.
 
 \begin{figure}[bht]
\begin{center}
\resizebox{3.15in}{!}{\includegraphics*[83,220][510,562]{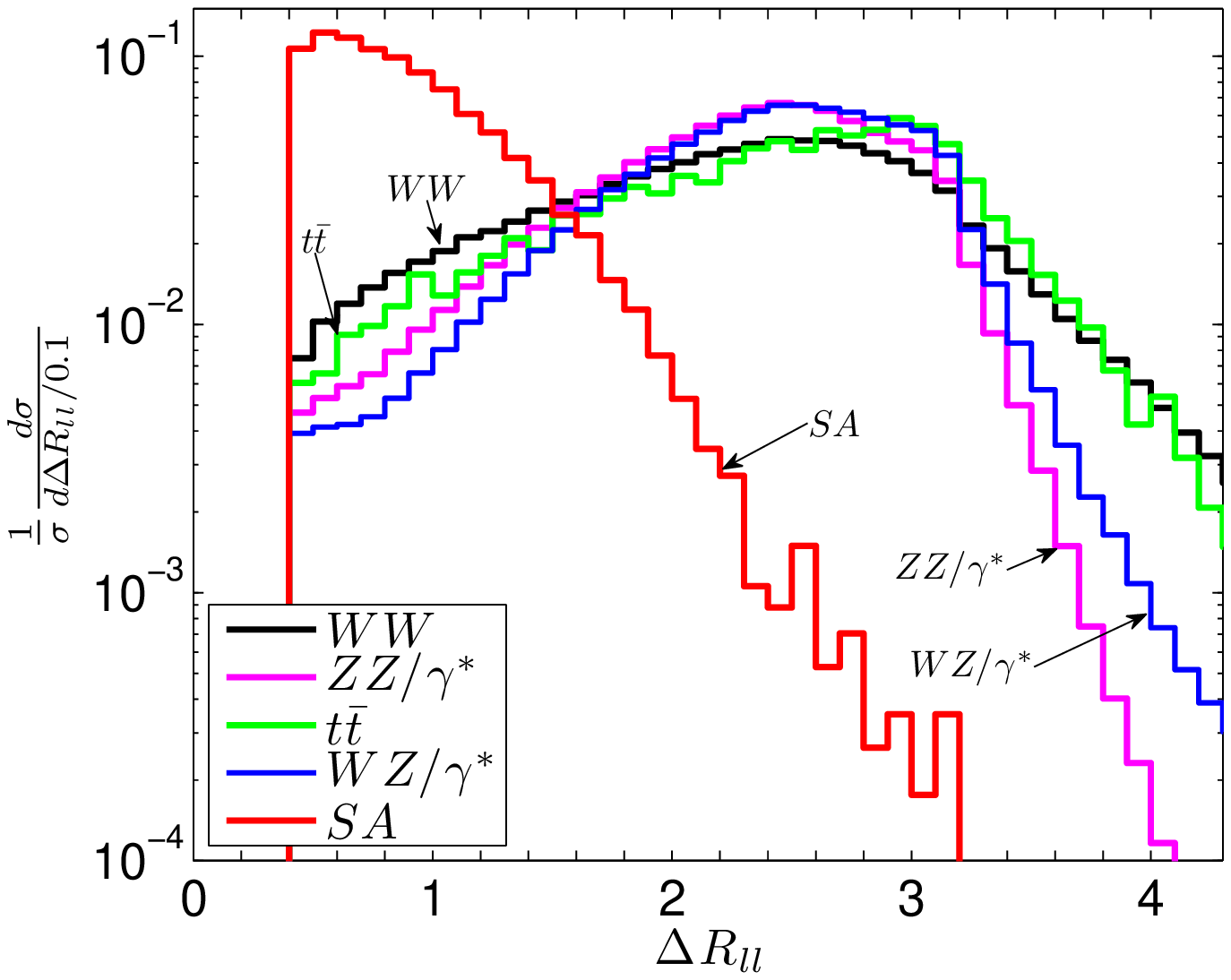}}
\resizebox{3.15in}{!}{\includegraphics*[83,220][510,562]{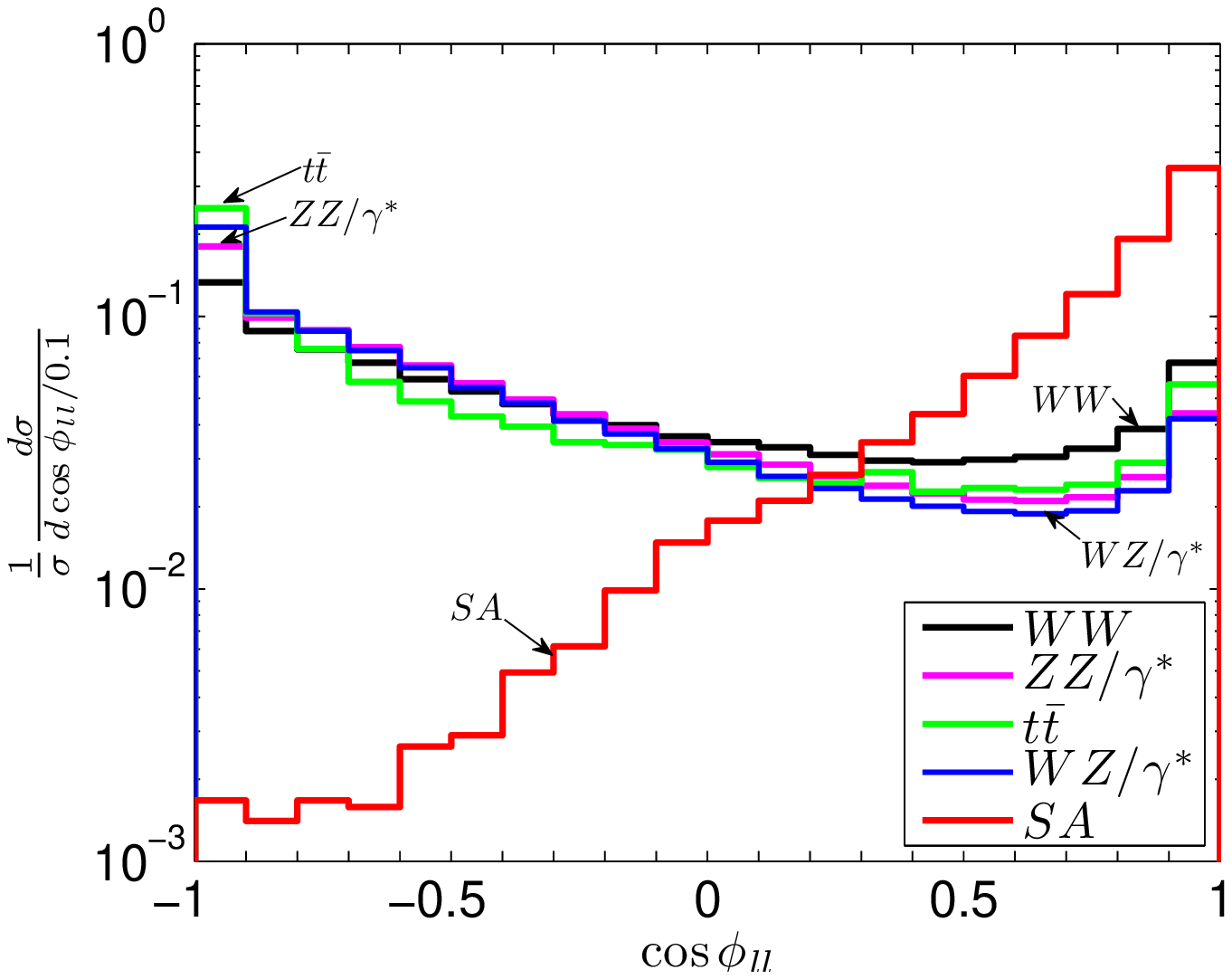}}
\caption{Distributions of the angular separation variables $\Delta R_{\ell \ell}$ 
(left panel) and $\cos\phi_{\ell \ell}$ (right panel) for benchmark point LH3, in
which decays of the pseudoscalar $A$ occur via an off-shell $Z$.  These 
distributions justify the imposition of the minimum $\cos\phi_{\ell\ell}$ and 
maximum $\Delta R_{\ell\ell}$ cuts described in the text.}
\label{fig:angle}
\end{center}
\end{figure}
 
To further suppress the Standard-Model $WW$, $Z\gamma^\ast$, $W\gamma^\ast$,
$Wt$, and $t\bar{t}$ backgrounds in cases in which $\delta_2 < M_Z$, it is useful 
to select events on the basis of observables related to
the angular separation between charged leptons.  The $\ell^+$ and $\ell^-$ 
produced by these SM background processes are typically energetic and 
well-separated from one another.  On the other hand, those resulting
from $A$ decay via an off-shell $Z$ tend to be soft, with small 
(and extremely so, if $\delta_2$ is quite small) angular separation.  This
difference in event topology is readily apparent from Fig.~\ref{fig:angle}, 
which displays the $\Delta R_{\ell\ell}$ (left panel) and 
$\cos\phi_{\ell\ell}\equiv \cos(\phi_{\ell^+}-\phi_{\ell^-})$ 
(right panel) distributions for benchmark LH3.  It is therefore useful,  
in cases in which $\delta_2 < M_Z$, to impose the additional cuts  
\begin{itemize}
\item $\Delta R_{\ell\ell}\leq \Delta R_{\ell\ell}^{\mathrm{max}}$,
\item $\cos\phi_{\ell\ell}\geq \cos\phi_{\ell\ell}^{\mathrm{min}}$,
\end{itemize}
where $\Delta R^{\mathrm{max}}$ and
$\cos\phi_{\ell\ell}^{\mathrm{min}}$ are to be optimized for
each benchmark point. 

In certain situations, the imposition of additional
event-selection criteria can also be helpful in distinguishing
signal from background events. 
For example, it can also be advantageous to impose a minimum 
cut on the total transverse momentum variable 
\begin{equation}
  H_T\equiv \displaystyle{\not}p_T +\sum_i p_T^{\ell_i},
\end{equation}  
which can serve as an efficient discriminant in both the 
$\delta_2>M_Z$ and $\delta_2<M_Z$ cases: 
\begin{itemize}
\item $H_T \geq H_T^{\mathrm{min}}$.
\end{itemize}
Again, the  threshold $H_{T}^{\mathrm{min}}$ 
can be optimized to suit a given benchmark point.  This
cut can be helpful in reducing the $WW$ and $Z\gamma^\ast$
backgrounds, but is less so in reducing the contribution from 
$t\bar{t}$.  In addition, it can sometimes also be useful to tighten 
the minimum $\displaystyle{\not}E_T$ cut applied during the
Level~II cuts.  Therefore, at Level~III, we allow for the imposition 
of an additional missing-transverse-energy cut
\begin{itemize}
\item $\displaystyle{\not}E_T \geq \displaystyle{\not}E_T^{\mathrm{min}}$.
\end{itemize}
Furthermore, in cases in which 
$\delta_2$ is small and the charged leptons associated with the signal 
process far less energetic than those associated with the SM backgrounds, 
it can be useful to impose a ceiling 
$p_{T\ell}^{\rm max}$ on the $p_T$ of each charged lepton, as we do
here for benchmark point LH5. 

\begin{itemize}
\item $p_{T\ell} \leq p_{T\ell}^{\mathrm{max}}$.
\end{itemize}
     
\begin{table}
\begin{center}
\begin{tabular}{|c|cc|cc|ccc|}\hline
~~Benchmark~~ &
  ~~~$M_{\ell\ell}^{\mathrm{min}}$~~~ & ~~~$M_{\ell\ell}^{\mathrm{max}}$~~~ & 
  ~~$\Delta R_{\ell\ell}^{\mathrm{max}}$~~ & 
  ~~$\cos\phi_{\ell\ell}^{\mathrm{min}}$~~& 
  ~~~~$H_T^{\mathrm{min}}$~~~~ & 
  ~~~~$\displaystyle{\not}E_T^{\mathrm{min}}$~~~~ &
  ~~~$p_{T\ell}^{\mathrm{max}}$~~~ \\\hline
LH1 & ~80~GeV~ & ~100~GeV~ & $-$ & $-$ & ~150~GeV~ &  ~50~GeV~ & $-$ \\
LH2 & $-$      &  ~70~GeV~ & 1.2 & 0.7 & ~200~GeV~ & ~100~GeV~ & $-$ \\
LH3 & ~20~GeV~ &  ~50~GeV~ & 0.8 & 0.7 & ~200~GeV~ &  ~90~GeV~ & $-$ \\
LH4 & ~20~GeV~ &  ~50~GeV~ & 0.8 & 0.7 & ~200~GeV~ &  ~90~GeV~ & $-$ \\
LH5 & $-$      &  ~10~GeV~ & 0.6 & 0.9 & $-$       &  ~30~GeV~ & ~25~GeV~ \\
\hline
HH1 & ~80~GeV~ & ~100~GeV~ & 2.0 & $-$ & ~200~GeV~ &  ~80~GeV~ & $-$ \\
HH2 & ~20~GeV~ &  ~70~GeV~ & 1.2 & 0.7 & ~200~GeV~ &  ~90~GeV~ & \\
HH3 & $-$      &  ~25~GeV~ & $-$ & $-$ & $-$       &  ~30~GeV~ & $-$ \\
\hline
\end{tabular}
\caption{A list of the optimized Level~III cuts used in the analysis of each of the benchmark  points presented in Table~\ref{tab:BMs}. An entry of ``$-$"  indicates that the corresponding 
cut is not imposed.  Note that a $\displaystyle{\not}E_T^{\rm min}$ cut of 30~GeV has been
applied in each of these scenarios as a part of the Level~II cuts, but that this threshold
has been raised for several of the points at Level~III.  For more details on the definition 
of the thresholds used, see text.\label{tab:CutParams}}
\end{center}
\end{table}
 
In Table~\ref{tab:CutParams}, we list the Level~III cuts applied in each 
of the benchmark scenarios listed in Table~\ref{tab:BMs}.
The precise numbers appearing in this table have been selected in order
to maximize the $S/\sqrt{B}$ ratio for each individual benchmark point.
It should be noted that the particular set of cuts applied in each case 
indeed depends primarily on the relationship between $\delta_2$ and $M_Z$. 


\section{Results\label{sec:Results}}

Now that we have discussed in detail the event-selection procedure
to be used in our numerical analysis of dilepton signals in the IDM, 
we turn to present the results of that numerical analysis.
In Table~\ref{tab:BMXSecscutIII}, we list the cross-sections for the 
signal process and the most relevant backgrounds after the application 
of our Level I+II+III cuts.  The last two columns in the Table display 
the signal-to-background ratio $S/B$ and the statistical 
significance (as given by $S/\sqrt{B}$ at an integrated luminosity of 
${\cal L}=100\mbox{ fb}^{-1}$) for each benchmark 
point\footnote{One modification is made in the case of LH5.  For this point,
both signal and background event rates are quite low, and consequently
the significance value quoted in the  last column of 
Table~\ref{tab:BMXSecscutIII} was obtained using Poisson statistics
rather than $S/\sqrt{B}$.} in our analysis,
after the implementation of these same cuts. 
Note that the numbers quoted here for benchmark LH4 with small 
$\delta_1$ include, in addition to the usual 
$pp\rightarrow SA\rightarrow \ell^+\ell^-+\displaystyle{\not}E_T$ contribution,
contributions from the processes 
$pp\rightarrow H^\pm A\rightarrow \ell^+\ell^-jj+\displaystyle{\not}E_T$ and
$pp\rightarrow H^\pm A\rightarrow \ell^+\ell^-\ell^\pm+\displaystyle{\not}E_T$
in which the additional jets or leptons from $H^\pm$ decay are sufficiently 
soft as to escape detection.  It should be noted that taking these
contributions into account results in an increase in the 
statistical significance of discovery in this channel from 
$2.07\sigma$ to $3.29\sigma$.  For the other benchmark points listed
in Table~\ref{tab:BMs}, $\delta_1\geq 50$~GeV, and consequently the
contribution from $pp\rightarrow H^\pm A$ processes with soft jets
or leptons will be negligible. 

\begin{table}
\begin{center}
\begin{tabular}{|c|c|cccccccc|cc|}\hline
 &\multicolumn{9}{c|}{Level~III Cuts}
&& \\ \cline{2-10}     
~Benchmark~ &~$\sigma_{SA}$ ~&~$\sigma_{H^+H^-}$ ~ &~$\sigma_{hZ}$ ~        
          &~$\sigma_{WW}$ ~&~$\sigma_{ZZ/\gamma^\ast}$ ~
          &~$\sigma_{t\bar{t}}$ ~&~$\sigma_{WZ/\gamma^\ast}$ ~ & ~$\sigma_{Wt}$ ~ 
          &~$\sigma_{\mathrm{BG}}^{\mathrm{comb}}$ ~&~~$S/B$~&~$S/\sqrt{B}$~ \\
&(fb)&(fb)&(fb)&(fb)&(fb)&(fb)&(fb)&(fb)&(fb)&&\\ \hline
LH1 & 3.42 &0.04    & 1.28   & 11.59& 36.99 &4.55    &19.52& 3.82 &77.79 & 0.04&  3.87 \\
LH2 & 0.89 &$\sim 0$& 0.01   &  0.07&  0.24 &0.11    & 0.08& 0.07 & 0.58 & 1.53& 11.66 \\
LH3 & 0.18 &$\sim 0$&$\sim 0$&  0.03&  0.15 &0.05    & 0.04& 0.06 & 0.34 & 0.52&  3.04 \\
LH4 & 0.19 &$\sim 0$&  0     &  0.03&  0.15 &0.05    & 0.04& 0.06 & 0.34 & 0.57&  3.29 \\
LH5 & 0.004&$\sim 0$&$\sim 0$&  0.13&  0.04 &$\sim 0$& 0.04& 0.01 & 0.23 & 0.02&  0.02 \\\hline
HH1 & 0.65 &$\sim 0$&  0     &  0.45& 13.41 &0.55    & 5.85& 0.45 &20.71 & 0.03&  1.42 \\
HH2 & 0.37 &0.01    &  0     &  0.08&  0.26 &0.12    & 0.09& 0.12 & 0.67 & 0.56&  4.55 \\
HH3 & 1.01 &$\sim 0$&  0     & 17.49&  1.06 &1.60    & 0.76& 1.65 &22.56 & 0.04&  2.12 \\\hline
\end{tabular}
\caption{Cross-sections for the processes 
$pp\rightarrow SA\rightarrow \ell^+\ell^-  \displaystyle{\not} E_T$, 
$pp\rightarrow H^+H^-\rightarrow \ell^+\ell^-  \displaystyle{\not} E_T$, and
$pp\rightarrow h^{(*)}Z\rightarrow \ell^+\ell^-  \displaystyle{\not} E_T$ 
at the LHC for each of the benchmark points presented in Table~\ref{tab:BMs}
after the application of our Level~III cuts. 
Cross-sections for the dominant SM backgrounds ($WW$, $ZZ/\gamma^*$, etc.) after the
application of the Level~III cuts are also shown, as is the total background 
cross-section including all of these individual contributions.  An 
entry of ``$\sim 0$'' indicates a cross-section less than $1$~ab.  
The last two columns display the signal-to-background ratio 
$S/B$ and statistical significance (as given by $S/\sqrt{B}$)
corresponding to an integrated luminosity of 
${\cal L}=100\mbox{ fb}^{-1}$ after the application of these same cuts.
\label{tab:BMXSecscutIII}}
\end{center}
\end{table}

Let us now turn to examine the results for each of the individual
benchmark scenarios in our study in more detail.  We begin with 
the LH1, which involves a light Higgs boson, a light LIP ($m_S\sim 40$~GeV), and 
a large mass splitting ($\delta_2=100$ GeV $>M_Z$).  The dominant backgrounds 
in this scenario are those from $ZZ$ and $WZ$, each of which also involves the 
leptonic decay of an on-shell $Z$ and is therefore difficult to differentiate 
from the signal process on the basis of kinematical variables.
The remaining backgrounds are efficiently suppressed 
after the imposition of the $M_{\ell\ell}$ cut near the $M_Z$ window.
With $100 \mbox{ fb}^{-1}$ of integrated 
luminosity, a significance level of $3.87\sigma$ could be obtained in this 
benchmark scenario.  The situation for the heavy-Higgs benchmark HH1 is similar; 
however the smaller $SA$-production cross-section in this case 
(due primarily to an increased LIP mass) translates into a lower statistical 
significance. 

Benchmark point LH2 also includes a $40$~GeV dark matter particle, 
but involves a smaller mass splitting than that of LH1: $\delta_2=70$~GeV.  This scenario 
affords the best opportunity for discovery at the LHC out of any of the benchmark 
points in our analysis, yielding a statistical significance of $11.66\sigma$ with
$100\mbox{ fb}^{-1}$ of integrated luminosity.  Two factors contribute to its success:  
a small production threshold $m_S+m_A$, and the fact that $\delta_2<M_Z$, which implies
that the $CP$-odd scalar $A$ decays   via an off-shell $Z$.  The latter consideration makes it possible
to eliminate $ZZ$ and $WZ$ background contributions quite efficiently by setting
$M_{\ell\ell}^{\mathrm{max}}$ comfortably below the $Z$ pole.  Further cuts on the
angular variables $\cos\phi_{\ell\ell}$ and $\Delta R_{\ell\ell}$ serve to reduce
the remaining backgrounds to a manageable level.  After all cuts are imposed, 
events from the low-$M_{\ell\ell}$ tail of the $ZZ$ distribution form the
dominant background.  It should be noted, however, that while the aforementioned    
angular-separation cuts are quite efficient in reducing background events, this
efficiency comes with a price: the cuts also eliminate a substantial fraction of 
signal events.  This explains why the signal cross-section for LH2 is less than 
that for LH1, as no angular separation cuts are imposed in the latter scenario.   

Benchmarks LH3 and LH4 are superficially similar, given that they involve a 
similar LIP mass $m_S\sim 80$~GeV and the same mass splitting $\delta_2=50$ GeV. 
In this case, however, the marked difference in $\delta_1$ --- a parameter 
which generally has little effect on observability of the dilepton signal in the  
$SA$ associated-production channel --- between the two points has a substantial
impact on their collider phenomenology.  The reason for this is twofold.  First of
all, since $\delta_2>\delta_1$ for LH4 (unlike any other benchmark in our analysis), 
the decay channels $A\rightarrow H^\pm W^\mp\rightarrow X +\displaystyle{\not}E_T$, 
where $X$ denotes either four jets, two jets and a single charged lepton, or two
charged leptons, are open in this scenario, with a branching ratio 
$\mathrm{BR}(A\rightarrow H^\pm W^\mp\rightarrow X +\displaystyle{\not}E_T)=0.435$.  
As a result, $\mathrm{BR}(A\rightarrow SZ\rightarrow \ell^+\ell^- +\displaystyle{\not}E_T)$,
and thus the dilepton signal cross-section, are reduced by an additional  
factor of two relative to
those points for which such competing decays are kinematically prohibited.
Second of all, as discussed above, the small value for $\delta_1=10$ GeV in LH4 
allows the additional contribution of $AH^\pm$ process to the signal due to the 
unobservable soft jets and leptons from $H^\pm$ decay.  These additional
contributions augment the overall signal cross-section and more than compensate
for the diminished 
$\mathrm{BR}(A\rightarrow SZ\rightarrow \ell^+\ell^-+\displaystyle{\not}E_T)$, as
discussed above.  For 
$100\mbox{ fb}^{-1}$ of integrated luminosity, a significance 
level greater than $3\sigma$ could be reached for LH4 as well as LH3. 

The final light-Higgs scenario in our analysis, LH5, turns out to be the most 
difficult benchmark point for which to observe a dilepton signal, 
primarily because of the small mass splitting $\delta_2=10$~GeV between $S$ and $A$.  
The charged leptons in the final state tend to be extremely soft, and consequently
the signal remains buried under the SM background even after an optimized set of
Level~III cuts is applied.  Scenarios with a small value of $\delta_2$ will in general 
be difficult to discover via this channel at the LHC. 
It should be noted that the results we obtain for this benchmark differ 
significantly from the parton-level results quoted in~\cite{CaoMa} 
for a similar benchmark scenario, also with $\delta_2=10$~GeV.  
The discrepancy owes primarily to our imposition of a Level I cut of
$\Delta R_{\ell\ell}>0.4$ cut designed to replicate the
effect of electron and muon isolation requirements 
at the ATLAS and CMS detectors.  Since the angular separation
between the lepton momenta tends to be extremely small for such
a small value of $\delta_2$, a vast majority of signal events
will have $\Delta R_{\ell\ell}<0.4$ and hence be eliminated by
this cut.  

Let us now turn to discuss the benchmark points which feature a heavy ($m_h=500$ GeV) 
Higgs boson --- in other words, those benchmarks for which the IDM successfully
addresses the LEP paradox.  While the electroweak precision constraints discussed 
in Sect.~\ref{sec:Constraints} are more stringent in this case, these constraints   
primarily affect $\delta_1$, which is typically required to be quite large.  Since
this parameter generally does not affect results in the dilepton channel, which
depend primarily on $m_S$ and $\delta_2$,   the same qualitative results obtained
for the light-Higgs benchmarks also apply here.   
For HH1, with $\delta_2=100$ GeV, a significance level of only $1.42\sigma$ can be achieved 
with $100\mbox{ fb}^{-1}$ of integrated luminosity, due to both the overwhelming SM 
backgrounds that exist for dilepton processes involving on-shell $Z$ decay, and a 
suppressed signal cross-section relative to benchmark LH1 (which has a far lighter LIP).
For HH3 --- a benchmark with a somewhat small value of $\delta_2$ --- a 
$M_{\ell\ell}^{\mathrm{max}}=25$~GeV cut helps to cut down the SM backgrounds from 
processes involving on-shell $Z$ decay.  It is, however, hard to improve upon the
statistical significance by implementing additional cuts.
The remaining background events
which survive this cut (most of which come from $WW$)   tend to have similar
$\cos\phi_{\ell\ell}$ and $\Delta R_{\ell\ell}$ distributions to those of the signal ---
a situation which makes the application of further, angular cuts essentially redundant.
Furthermore, since the missing-energy distribution for the signal events in scenarios 
with small $\delta_2$ peaks at relatively low values of $\displaystyle{\not}E_T$, there is 
little to be gained by increasing $\displaystyle{\not}E_T^{\mathrm{min}}$ much beyond the Level~II threshold
of $30$~GeV.  By contrast, in scenarios with larger $\delta_2$, an elevated missing-energy
cut works quite effectively in tandem with the angular cuts in reducing backgrounds from
$WW$ and $t\bar{t}$.  A significance level of $2.32\sigma$  is reached for HH3 with $100\mbox{ fb}^{-1}$ of integrated luminosity.  

It is benchmark HH2, however, which 
affords the best opportunity for discovery at the LHC from among the heavy-Higgs 
scenarios, with a statistical significance of $4.55\sigma$ at $100\mbox{ fb}^{-1}$ of 
integrated luminosity.  
This is because the signal for this benchmark can be distinguished from the $WZ$ 
and $ZZ$ backgrounds on the basis of $M_{\ell\ell}$ cuts, and from the remaining 
$WW$, $Wt$, and $t\bar{t}$ backgrounds on the basis of $\cos\phi_{\ell\ell}$, 
$\Delta R_{\ell\ell}$, and $\displaystyle{\not}E_T$ cuts in the same manner as for 
the low-Higgs-mass point LH2.  We therefore conclude that
even scenarios in which the IDM permits an evasion of the LEP upper bound on $m_h$ can
yield an observable dilepton signal at the LHC.

From the results in Table~\ref{tab:BMXSecscutIII}, it is evident that
the prospects for detecting a signal in the dilepton channel
in the IDM model hinge primarily on two criteria.  
The first of these is the dependence of the
cross-section for $q\bar{q}\rightarrow SA$ on $m_S+m_A$ and $\delta_2$.
This cross-section is, of course, larger in cases where the pair-production 
threshold energy $m_S+m_A$ is small.   Among cases with similar values of 
$m_S+m_A$, those in which $\delta_2$ is smaller will have larger
production cross-sections.  This can be understood by noting that the
partonic cross-section for this process depends on $m_S$ and $\delta_2$ in
the following way:
\begin{eqnarray}
  \hat{\sigma}_{q\bar{q}\rightarrow SA}(\hat{s}) & \propto &
 [\hat{s}^2-2\hat{s}(\delta_2(\delta_2+2m_S)+2m_S^2)+
       \delta_2^2(\delta_2+2m_S)^2]^{3/2}.
  \label{eq:PartonicXSec}
\end{eqnarray} 
For values of $\hat{s}\sim m_S + m_A$, for which the dependence of
this expression on $m_S$ and $\delta_2$ is non-negligible, it is apparent
that for fixed $m_S + m_A$, $\hat{\sigma}_{q\bar{q}\rightarrow SA}(\hat{s})$
decreases with increasing $\delta_2$.  This accounts for the difference 
between the $pp\rightarrow SA$ production cross-sections for benchmarks
LH1 and HH3 quoted in Table~\ref{tab:BMXSecs}. 

The second criterion is the relationship between $\delta_2$ and 
$M_Z$: cases in which $\delta_2<M_Z$ tend to have a higher
statistical significance than those in which $\delta_2>M_Z$, as is
manifest from comparing the results for benchmarks LH2 and LH1 in
Table~\ref{tab:BMXSecscutIII}.
This is because in the latter case, it is difficult to
distinguish the signal process from the dominant $ZZ$ background
on the basis of event topology.  On the other hand, when
$\delta_2$ is exceedingly small (as it is in our LH5 scenario),
the charged leptons will be so soft that the detector-acceptance 
(i.e.\ Level~I) cuts will eliminate the vast majority of would-be 
signal events, as discussed above.  Between these extremes, a window of
\begin{equation}
40\mbox{~GeV}\lesssim \delta_2\lesssim 80\mbox{~GeV}
\label{eq:OptDel2Range}
\end{equation}
emerges within which the prospects for observing a signal 
are quite good, so long as the LIP mass also falls roughly within the 
$40 - 80$~GeV range.  For cases in which $\delta_2 \gtrsim M_Z$, the
prospects for discovery at the LHC are reasonable --- meaning a statistical
significance around the $3\sigma$ level with 100 ${\rm fb}^{-1}$ of
integrated luminosity --- only if the dark-matter particle 
is light ($m_S\sim 40$~GeV).

It is not difficult simultaneously to satisfy the  
constraints discussed in Sect.~\ref{sec:Constraints} and
to realize a $\delta_2$ value within this mass window of $40 - 80$~GeV while keeping 
the LIP mass relatively light ($m_S\lesssim 80$ GeV) --- or, alternatively, to obtain 
a large mass splitting $\delta_2\gtrsim M_Z$ and a light LIP mass of around 40~GeV.
This is true not only in models where the Higgs boson is light and
the parameters of the theory comparatively unconstrained, but
also in cases in which the mechanism of 
Ref.~\cite{BarbieriIDM} for evading electroweak precision bounds on the
Higgs mass is realized in nature, and $m_h\sim 500$~GeV.  
In either case, it would be possible to observe a dilepton signal at the 
LHC at a significance level of $3\sigma$ or higher, with an integrated 
luminosity of $100\mbox{ fb}^{-1}$.


\section{Conclusion\label{sec:Conclusion}}

In this work, we have investigated the potential for observing a dilepton
signature in the Inert Doublet Model at the LHC.  We have explored the prospects
for a number of benchmark scenarios, including several in 
which the IDM successfully ameliorates the LEP paradox and the Higgs-boson
mass can be elevated as high as $m_h=400-500$~GeV, as well as several of 
the dark-matter motivated scenarios cataloged in Ref.~\cite{ArizonansDarkIDM}.   
We have shown that for cases in which the dark matter candidate is relatively light 
($40 - 80$~GeV) and $40 {\rm\ GeV} \lesssim \delta_2 \lesssim 80$ GeV, a signal with a significance of 
more than $3\sigma$ should be apparent at the LHC with less than $100\mbox{ fb}^{-1}$ 
of integrated luminosity.  Moreover, in cases when the LIP is on the lighter end of
this range, a $3\sigma$ discovery would be possible with only $10\mbox{ fb}^{-1}$ of 
integrated luminosity.  In addition, there are also certain cases in which 
$\delta_2>M_Z$ and the LIP is light ($m_S\sim 40$~GeV) for which the prospects 
for detection are also reasonably good.  

Of course the observation of an excess in the $\ell^+\ell^-+\displaystyle{\not}E_T$ 
channel alone, while exciting, is by no means
conclusive evidence for the Inert Doublet Model.  Indeed, many
models of beyond-the-Standard-Model physics lead to such a signature,
including weak-scale supersymmetry, two-Higgs-doublet models, etc.
Fortunately, evidence for the IDM can come from a number of other 
sources.  Some of these sources involve other channels associated 
with the SM-like Higgs  at the LHC.
One potentially interesting signal could arise  
due to deviations of the decay properties of the Higgs
boson $h$ from those of a SM Higgs.  In situations in which $m_h > 2 m_S$,
for example, $\Gamma(h\rightarrow SS)$ can contribute substantially  
to the invisible Higgs width.  Searches for the Weak-Boson Fusion (WBF)
process $qq'\rightarrow qq'h$, with $h$ decaying invisibly, can be used
effectively to identify a Higgs boson at the LHC~\cite{InvisibleHiggs},
and preliminary studies~\cite{CaoMa} indicate that a $5\sigma$ discovery
should be possible with only $10\mbox{ fb}^{-1}$ of integrated luminosity
in regions of parameter space where $\mathrm{BR}(h\rightarrow {\rm invisible})$ 
is large.
Moreover, if $m_h >2 m_A$,  the  tetralepton + $\displaystyle{\not} E_T$
signatures resulting from decays of the form 
$h\rightarrow AA\rightarrow SS\ell^+\ell^-\ell^+\ell^-$
may also be detectable in certain regions of parameter space.  The
observation of signals
of this sort, along with the non-observation of other signals which 
appear in standard 2HDM due to $\phi_i\bar{f}f'$ couplings (where
$\phi_i={H^\pm,A,S}$ and $f$ and $f'$ are SM fermions) absent in  
the Inert Doublet Model, could together serve to distinguish the IDM
from other scenarios for physics beyond the Standard Model.  
 
Evidence for the IDM could also come from a a variety of other 
sources, including dark-matter-direct-detection experiments and from the
observation of energetic 
gamma-rays~\cite{SwedesGammaRayIDM,ArizonansGammaRayIDM} 
or neutrinos~\cite{NeutrinoDarkIDM} resulting from 
LIP dark matter annihilation.  Clearly, the particular set of signals that an 
inert doublet would manifest differs substantially, depending 
on which of the allowed regions of parameter space the model 
happened to inhabit, and as we have shown, the 
$\ell^+\ell^- + \displaystyle{\not}E_T$ channel can provide an 
important probe into which region that might be.        


\section{Acknowledgments}

We would like to thank J. Alwall and T. Han for useful discussions and comments.
We would like to thank the Aspen Center of Physics, where part of this work 
was completed.  This work was supported in part
by the Department of Energy under Grant~DE-FG02-04ER-41298.


\smallskip


\end{document}